\documentclass[english,aps,manuscript,preprintnumbers]{revtex4-1}
\usepackage[T1]{fontenc}
\usepackage[latin9]{inputenc}
\usepackage{amssymb}
\usepackage{graphicx}
\usepackage{wasysym}
\usepackage{esint}
\usepackage{babel}
\begin{document}

\title{On decoherence in quantum gravity }

\author{Dmitriy \surname{Podolskiy}}
\email{Dmitriy_Podolskiy@hms.harvard.edu}

\affiliation{Harvard Medical School, 77 Avenue Louis Pasteur, Boston, MA, 02115}

\author{Robert \surname{Lanza}}

\affiliation{Wake Forest University, 1834 Wake Forest Rd., Winston-Salem, NC,
27106}

\date{\today}
\begin{abstract}
It was previously argued that the phenomenon of quantum gravitational
decoherence described by the Wheeler-DeWitt equation is responsible
for the emergence of the arrow of time. Here we show that the characteristic
spatio-temporal scales of quantum gravitational decoherence are typically
logarithmically larger than a characteristic curvature radius $R^{-1/2}$
of the background space-time. This largeness is a direct consequence
of the fact that gravity is a non-renormalizable theory, and the corresponding
effective field theory is nearly decoupled from matter degrees of
freedom in the physical limit $M_{P}\to\infty$. Therefore, as such,
quantum gravitational decoherence is too ineffective to guarantee
the emergence of the arrow of time and the ``quantum-to-classical''
transition to happen at scales of physical interest. We argue that
the emergence of the arrow of time is directly related to the nature
and properties of physical observer.
\end{abstract}
\maketitle

\section{Introduction \label{sec:Introduction}}

Quantum mechanical decoherence is one of the cornerstones of the quantum
theory \citep{Zurek2003,Joos2003}. Macroscopic physical systems are
known to decohere during vanishingly tiny fractions of a second, which,
as generally accepted, effectively leads to emergence of a deterministic
quasi-classical world which we experience. The theory of decoherence
has passed extensive experimental tests, and dynamics of the decoherence
process itself was many times observed in the laboratory \citep{Brune1996,Andrews1997,Arndt1999,Friedman2000,VanderWal2000,Kielpinski2001,Vion2002,Chiorescu2003,Hackermuller2003,Hackermuller2004,Martinis2005,Petta2005,Deleglise2008}.
The analysis of decoherence in non-relativistic quantum mechanical
systems is apparently based on the\emph{ notion of time}, the latter
itself believed to emerge due to decoherence between different WKB
branches of the solutions of the Wheeler-DeWitt equation describing
quantum gravity \citep{Zeh1989,Kiefer1992,Joos2003,Anastopoulos2013,Hu2014}.
Thus, to claim understanding of decoherence ``at large'', one has
to first understand decoherence in quantum gravity. The latter is
clearly problematic, as no consistent and complete theory of quantum
gravity has emerged yet.

Although it is generally believed that when describing dynamics of
decoherence in relativistic field theories and gravity one does not
face any fundamental difficulties and gravity decoheres quickly due
to interaction with matter \citep{Joos1986,Calzetta1995,Lombardo1996,Koksma2012},
we shall demonstrate here by simple estimates that decoherence of
quantum gravitational degrees of freedom might in some relevant cases
(in particular, in a physical situation realized in the very early
Universe) actually be rather ineffective. The nature of this ineffectiveness
is to a large degree related to the non-renormalizability of gravity.
To understand how the latter influences the dynamics of decoherence,
one can consider theories with a Landau pole such as the $\lambda\phi^{4}$
scalar field theory in $d=4$ dimensions. This theory is believed
to be trivial \citep{Aizenman1983}, since the physical coupling $\lambda_{{\rm phys}}$
vanishes in the continuum limit \footnote{There exist counter-arguments in favor of the existence of a genuine
strong coupling limit for $d=4.$\citep{Podolsky2010}}. When $d\ge5$, where the triviality is certain \citep{Aizenman1981,Aizenman1982},
critical exponents of $\lambda\phi^{4}$ theory and other theories
from the same universality class coincide with the ones predicted
by the mean field theory. Thus, such theories are effectively free
in the continuum limit, i.e., $\lambda_{{\rm phys}}\sim\frac{\lambda}{\Lambda^{d-4}}\to0$
when the UV cutoff $\Lambda\to\infty$. Quantum mechanical decoherence
of the field states in such QFTs can only proceed through the interaction
with other degrees of freedom. If such degrees of freedom are not
in the menu, decoherence is not simply slow, it is essentially absent. 

In effective field theory formulation of gravity dimensionless couplings
are suppressed by negative powers of the Planck mass $M_{P}$, which
plays the role of UV cutoff and becomes infinite in the decoupling
limit $M_{P}\to\infty$. Decoherence times for arbitrary configurations
of quantum gravitational degrees of freedom also grow with growing
$M_{P}$ although, as we shall see below, only logarithmically slowly
and become infinite at complete decoupling. If we recall that gravity
is \emph{almost} decoupled from physical matter in the real physical
world, ineffectiveness of quantum gravitational decoherence does not
seem any longer so surprising. While matter degrees of freedom propagating
on a fixed or slightly perturbed background space-time corresponding
to a fixed solution branch of the WdW equation decohere very rapidly,
decoherence of different WKB solution branches remains a question
from the realm of quantum gravity. Thus, we would like to argue that
in order to fit the ineffectiveness of quantum gravitational decoherence
and a nearly perfectly decohered world which we experience in experiments,
some additional physical arguments are necessary based on properties
of observer, in particular, her/his ability to process and remember
information.

This paper is organized as follows. We discuss decoherence in non-renormalizable
quantum field theories and relation between non-renormalizable QFTs
and classical statistical systems with first order phase transition
in Section \ref{sec:nonrenorm-QFTs}. We discuss decoherence in non-renormalizable
field theories in Section \ref{sec:Decoherence-non-renormalizable}
using both first- and second-quantized formalisms. Section \ref{sec:Decoherence-in-dS}
is devoted to the discussion of decoherence in dS space-time. We also
argue that meta-observers in dS space-time should not be expected
to experience effects of decoherence. Standard approaches to quantum
gravitational decoherence based on analysis of WdW solutions and master
equation for the density matrix of quantum gravitational degrees of
freedom are reviewed in Section \ref{sec:DecoherenceQG}. Finally,
we argue in Section \ref{sec:Conclusions} that one of the mechanisms
responsible for the emergence of the arrow of time is related to ability
of observers to preserve information about experienced events. 

\section{Preliminary notes on non-renormalizable field theories\label{sec:nonrenorm-QFTs}}

To develop a quantitative approach for studying decoherence in non-renormalizable
field theories, it is instructive to use the duality between quantum
field theories in $d$ space-time dimensions and statistical physics
models in $d$ spatial dimensions. In other words, to gain some intuition
regarding behavior of non-renormalizable quantum field theories, one
can first analyze the behavior of their statistical physis counterparts
describing behavior of classical systems with appropriate symmetries
near the phase transition. 

Consider for example a large class of non-renormalizable QFTs, which
includes theories with \emph{global} discrete and continuous symmetries
in the number of space-time dimensions higher than the upper critical
dimension $d_{{\rm up}}$: $d>d_{{\rm up}}$. Euclidean versions of
such theories are known to describe a vicinity of the 1st order phase
transition on the lattice \citep{A.M.Polyakov1987}, and their continuum
limits do not formally exist \footnote{Similarly, Euclidean $Z_{2}$, $O(2)$ and $SU(N)$ gauge field theories
all known to possess a first order phase transition on the lattice
at $d>d_{{\rm up}}=4$. }: even at close proximity of the critical temperature $T=T_{c}$ physical
correlation length of the theory $\xi\sim m_{{\rm phys}}^{-1}\sim(T-T_{c})^{-1/2}$
never becomes infinite. 

One notable example of such a theory is the $\lambda(\phi^{2}-v^{2})^{2}$
scalar statistical field theory, describing behavior of the order
parameter $\phi$ in the nearly critial system with discrete $Z_{2}$
symmetry. This theory is trivial \citep{Aizenman1981,Aizenman1982}
in $d>d_{{\rm up}}=4$ \footnote{Most probably, it is trivial even in $d=4$ \citep{Aizenman1983},
where it features a Landau pole (although there exist arguments in
favor of a non-trivial behavior at strong coupling, see for example
\citep{Podolsky2010}).}. Triviality roughly follows from the observation that the effective
dimensionless coupling falls off as $\lambda/\xi^{d-4}$, when the
continuum limit $\xi\to\infty$ is approached. 

\begin{figure}

\includegraphics{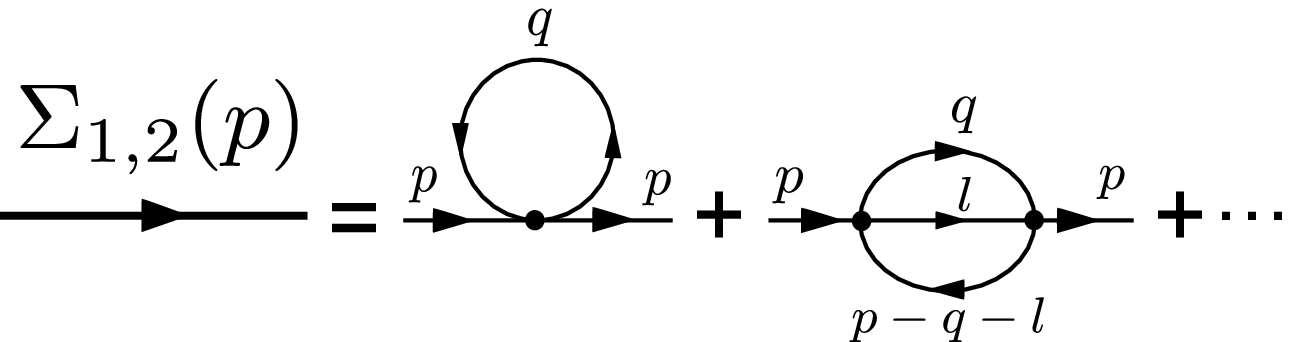}\caption{\label{fig:One-and-two-loop}One- and two-loop contributions to $\Sigma(p)$
in $\lambda\phi^{4}$ EFT.}

\end{figure}

What does it mean physically? First, the behavior of the theory in
$d>4$ is well approximated by mean field. This can be readily seen
when applying Ginzburg criterion for the applicability of mean field
approximation \citep{Cardy1996}: at $d>4$ the mean field theory
description is applicable arbitrarily close to the critical temperature.
This is also easy to check at the diagrammatic level: the two-point
function of the field $\phi$ has the following form in momentum representation
\[
\langle\phi(-p)\phi(p)\rangle\sim(p^{2}+m_{0}^{2}+\Sigma(p))^{-1},
\]
where $m_{{\rm 0}}^{2}=a(T-T_{c})$, and at one loop level (see Fig.
\ref{fig:One-and-two-loop})
\begin{equation}
\Sigma(p)\sim c_{1}g\Lambda^{2}+c_{2}g\Lambda^{2}\left(\frac{a(T-T_{c})}{\Lambda^{2}}\right)^{d/2-1},\label{eq:phi4Sigma}
\end{equation}
where $g=\lambda\Lambda^{d-4}$ is the dimensionless coupling. The
first term in the r.h.s. of (\ref{eq:phi4Sigma}) represents the mean
field correction leading to the renormalization/redefinition of $T_{c}$.
The second term is strongly suppressed at $d>4$ in comparison to
the first one. The same applies to any high order corrections in powers
of $\lambda$ as well as corrections from any other local terms $\sim\phi^{6},\phi^{8},\ldots,p^{m}\phi^{n},\ldots$
in the effective Lagrangian of the theory. 

As we see, the behavior of the theory is in fact simple despite its
non-renormalizability; naively, since the coupling constant $\lambda$
has a dimension $[l]^{d-4}$, one expects uncontrollable power-law
corrections to observables and coupling constants of the theory. Nevertheless,
as (\ref{eq:phi4Sigma}) implies, the perturbation theory series can
be re-summed in such a way that only mean field terms survive. Physics-wise,
it is also clear why one comes to this conclusion. At $d>4$ $Z_{2}$-invariant
statistical physics models do not possess a second order phase transition,
but of course do possess a first order one \footnote{This is equivalent to the statement that trvial theories do not admit
continuum limit.}. Behavior of the theory in the vicinity of the first order phase
transition can always be described in the mean field approximation,
in terms of the homogeneous order parameter $\Phi=\langle\phi\rangle$. 

\begin{figure}

\includegraphics[scale=0.8]{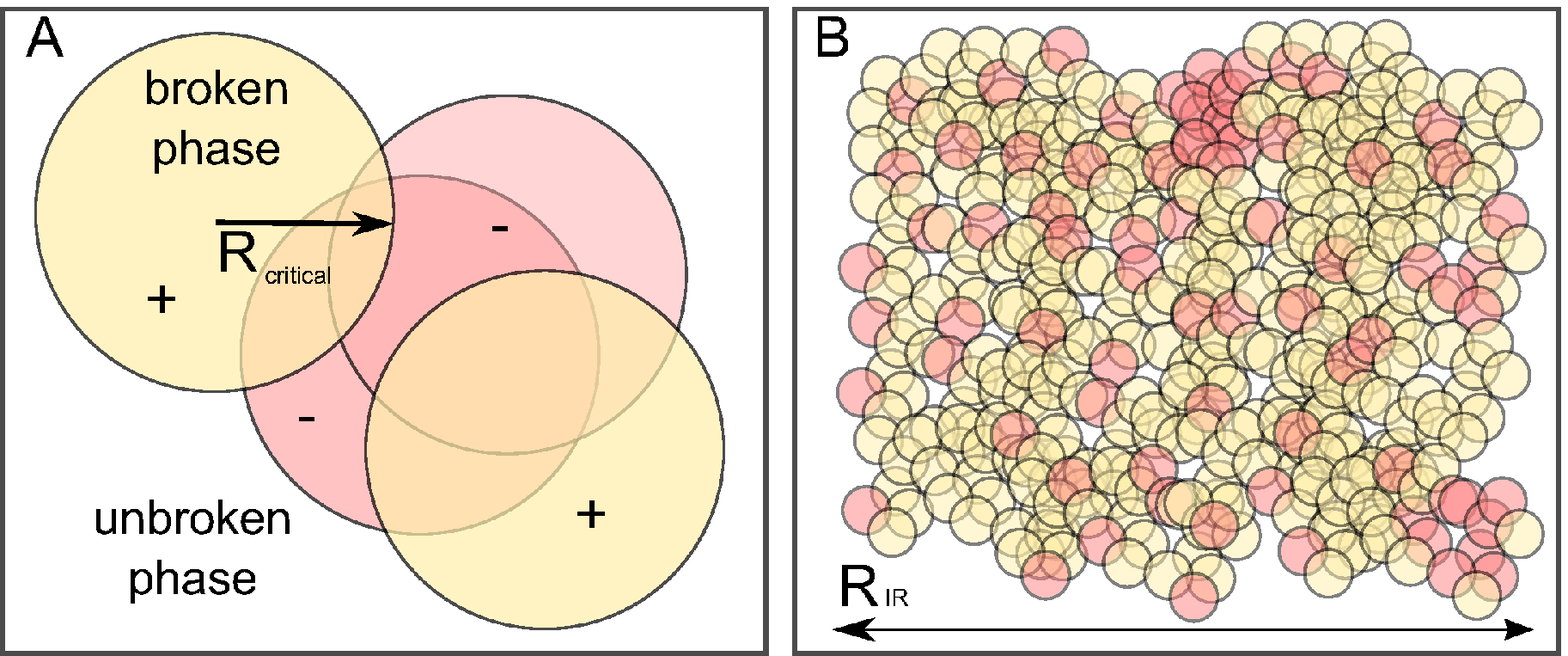}\caption{A possible configuration of order parameter in the $Z_{2}$ statistial
model in $d\ge5$ spatial dimensions. The left panel represents the
configuration of the field at scales slightly larger than the critical
radius $R_{{\rm crit}}\sim\xi$, that coinsides with the size of bubbles
of the true vacuum with broken $Z_{2}$ symmetry; $+$ corresponds
to bubbles with the vacuum $+\Phi_{0}$ inside, and $-$ - to the
bubbles with the vacuum $-\Phi_{0}$. At much larger scales of the
order of $R_{{\rm IR}}$ given by the expression (\ref{eq:phi4IRscale})
$\langle\Phi\rangle=0$ in average, as the contribution of multiple
bubbles with $\Phi=+\Phi_{0}$ is compensated by the contribution
of bubbles with $\Phi=-\Phi_{0}$. \label{fig:StatPhysConfiguration}}

\end{figure}

Our argument is not entirely complete as there is a minor culprit.
Assume that an effective field theory with the EFT cutoff $\Lambda$
coinciding with the physical cutoff is considered. Near the point
of the 1st order phase transition, when the very small spatial scales
(much smaller than the correlation length $\xi$ of the theory) are
probed, it is almost guaranteed that the probed physics is the one
of the broken phase. The first order phase transition proceeds through
the nucleation of bubbles of a critical size $R\sim(T-T_{c})^{-1/2}\sim\xi$,
thus very small scales correspond to physics inside a bubble of the
true vacuum $\langle\phi\rangle=\pm v$, and the EFT of the field
$\delta\phi=\phi-\langle\phi\rangle$ is a good description of the
behavior of the theory at such scales. As the spatial probe scale
increases, such description will inevitably break down at the IR scale
\begin{equation}
R_{IR}\sim m^{-1}\exp\left(\frac{{\rm Const.}}{\lambda m^{d-4}}\right)\sim\frac{\Lambda^{\frac{d-4}{2}}}{\sqrt{g}v}\exp\left(\frac{{\rm Const.}\Lambda^{(d-4)(d/2-1)}}{g^{d/2-1}v^{d-4}}\right),\label{eq:phi4IRscale}
\end{equation}
where $m\sim\xi^{-1}\sim\sqrt{\lambda}v\to0$ in the pre-critical
limit. This scale is directly related to the nucleation rate of bubbles:
at scales much larger than the bubble size $R$ one has to take into
account the stochastic background of the ensemble of bubbles of true
vacuum on top of the false vacuum, and deviation of it from the the
single-bubble background $\langle\phi\rangle=\pm v$ leads to the
breakdown of the effective field theory description, see Fig. \ref{fig:StatPhysConfiguration}.
Spatial homogeneity is also broken at scales $m^{-1}<l\ll R_{{\rm IR}}$
by this stochastic background, and this large-scale spatial inhomogeneity
is one of the reasons of the EFT description breakdown. 

Finally, if the probe scale is much larger than $\xi\sim(T-T_{c})^{-1/2}$
(say, roughly, of the order of $R_{IR}$ or larger), the observer
probes a false vacuum phase with $\langle\phi\rangle=0$. $Z_{2}$
symmetry dictates the existence of two true minima $\langle\phi\rangle=\pm v$,
and different bubbles have different vacua among the two realized
inside them. If one waits long enough, the process of constant bubble
nucleation will lead to self-averaging of the observed $\langle\phi\rangle$.
As a result, the ``true'' $\langle\phi\rangle$ measured over very
long spatial scales is always zero.

The main conclusion of this Section is that despite the EFT breakdown
at both UV (momenta $p\gtrsim\Lambda$) and IR (momenta $p\lesssim R_{IR}^{-1}$
) scales, the non-renormalizable statistical $\lambda\phi^{4}$ theory
perfectly remains under control: one can effectively use a description
in terms of EFT at small scales $R_{IR}^{-1}\lesssim p\lesssim\Lambda$
and a mean field at large scales. In all cases, the physical system
remains nearly completely described in terms of the homogeneous order
parameter $\Phi=\langle\phi\rangle$ or a ``master field'', as its
fluctuations are almost decoupled. Let us now see what this conclusion
means for the quantum counterparts of the discussed statistical physics
systems.

\section{Decoherence in relativistic non-renormalizable field theories\label{sec:Decoherence-non-renormalizable}}

We first focus on the quantum field theory with global $Z_{2}$-symmetry.
All of the above (possibility of EFT descriptions at both $R_{IR}^{-1}\lesssim E\lesssim\Lambda$
and $E\ll R_{IR}^{-1}$, breakdown of EFT at $E\sim\Lambda$ and $E\sim R_{IR}^{-1}$
with $R_{IR}$ given by the expression (\ref{eq:phi4IRscale})) can
be applied to the quantum theory, but there is an important addition
concerning decoherence, which we shall now discuss in more details.

\subsection{Master field and fluctuations\label{subsec:Master-field}}

As we discussed above, for the partition function of the $Z_{2}-$invariant
statistical field theory describing a vicinity of a first order phase
transition $\frac{T-T_{c}}{T_{c}}\ll1$ one approximately has 
\begin{equation}
Z=\int{\cal D}\phi\exp\left(-\int d^{d}x\left(\frac{1}{2}(\partial\phi)^{2}\pm\frac{1}{2}m^{2}\phi^{2}+\frac{1}{4}\lambda\phi^{4}+\ldots\right)\right)\approx\label{eq:PartitionFunction1}
\end{equation}
\begin{equation}
\approx\int d\Phi\exp\left(\mp\frac{1}{2}V_{d}m^{2}\Phi^{2}-\frac{1}{4}V_{d}\lambda\Phi^{4}-V_{d}\mu\Phi\right),\label{eq:PartitionFunction2}
\end{equation}
where $V_{d}$ is the $d-$volume of the system, and $d\ge5$ as in
the previous Section. Physically, the spatial fluctuations of the
order parameter $\phi$ are suppressed, and the system is well described
by statistical properties of the homogeneous order parameter $\Phi\sim\langle\phi\rangle$. 

The Wick rotated quantum counterpart of the statistical physics model
(\ref{eq:PartitionFunction1}) is determined by the expression for
the quantum mechanical ``amplitude''
\[
A(\Phi_{0},t_{0};\Phi,t)\approx\int d\Phi\exp\left(iV_{d-1}T\left(\mp\frac{1}{2}m^{2}\Phi^{2}-\frac{1}{4}\lambda\Phi^{4}\right)\right)=
\]
\begin{equation}
=\int_{\Phi(t_{0})=\Phi_{0}}^{\Phi(t)=\Phi}{\cal D}\Phi\exp\left(iV_{d-1}\int_{t_{0}}^{t}dt\left(\mp\frac{1}{2}m^{2}\Phi^{2}-\frac{1}{4}\lambda\Phi^{4}\right)\right),\label{eq:Amplitude}
\end{equation}
written entirely in terms of the ``master field'' $\Phi$ (as usual,
$V_{d-1}=\int d^{d-1}x$ is the volume of $(d-1)$-dimensional space).
In other words, in the first approximation the non-renormalizable
$\lambda\phi^{4}$ theory in $d\ge5$ dimensions can be described
in terms of a master field $\Phi$, roughly homogeneous in space-time.
As usual, the wave function of the field can be described as
\[
\Psi(\Phi,t)\sim A(\Phi_{0},t_{0};\Phi,t),
\]
where $\Phi_{0}$ and $t_{0}$ are fixed, while $\Phi$ and $t$ are
varied, and the density matrix is given by
\begin{equation}
\rho(\Phi,\Phi',t)={\rm Tr}\Psi(\Phi,t)\Psi^{*}(\Phi',t),\label{eq:DensityMatrixNREN}
\end{equation}
where the trace is taken over the degrees of freedom not included
into $\Phi$ and $\Phi'$, namely, fluctuations of the field $\delta\phi$
above the master field configuration $\Phi$. The contribution of
the latter can be described using the prescription 
\[
A\sim\int d\Phi{\cal D}\delta\phi\exp(iV_{d-1}T\left(\mp\frac{1}{2}m^{2}\Phi^{2}-\frac{1}{4}\lambda\Phi^{4}\right))\times
\]
\begin{equation}
\times\exp\left(i\int d^{d}x\left(\frac{1}{2}(\partial\delta\phi)^{2}\mp\frac{1}{2}m^{2}\delta\phi^{2}-\frac{3}{2}\lambda\Phi^{2}\delta\phi^{2}-\lambda\Phi^{3}\delta\phi-\lambda\Phi\delta\phi^{3}-\ldots\right)\right).\label{eq:AmplitudeAndFluctuations}
\end{equation}
In the ``mean field'' approximation (corresponding to the continuum
limit) $\lambda\to0$ fluctuations $\delta\phi$ are completely decoupled
from the master field $\Phi$, making (\ref{eq:Amplitude}) a good
approximation of the theory. To conclude, one physical consequence
of the triviality of statistical physics models describing vicinity
of a first order phase transition is that in their quantum counterparts
decoherence of entangled states of the master field $\Phi$ does not
proceed. 

\subsection{Decoherence in the EFT picture\label{subsec:Decoherence-in-EFT}}

When the correlation length $\xi\sim m_{{\rm phys}}^{-1}$ is large
but finite, decoherence takes a finite but large amount of time, essentially,
as we shall see, determined by the magnitude of $\xi$. This time
scale will now be estimated by two different methods.

As non-renormalizable QFTs admit an EFT description (which eventually
breaks down), dynamics of decoherence in such theories strongly depends
on the probe scale, coarse-graining effectively performed by the observer.
Consider a spatio-temporal coarse-graining scale $l>\Lambda^{-1}$
and assume that all modes of the field $\phi$ with energies/momenta
$l^{-1}<p\ll\Lambda$ represent the ``environment'', and interaction
with them leads to the decoherence of the observed modes with momenta
$p<l^{-1}$. If also $p>R_{IR}^{-1}$, EFT expansion near $\langle\phi\rangle$
is applicable. In practice, similar to Kenneth Wilson's prescription
for renormalization group analysis, we separate the field $\phi$
into the fast, $\phi_{f}$, and slow, $\phi_{s}$, components, considering
$\phi_{f}$ as an environment, and since translational invariance
holds ``at large'', $\phi_{s}$ and $\phi_{f}$ are linearly separable
\footnote{A note should be taken at this point regarding the momentum representation
of the modes. As usual, $\phi_{f}$ is defined as integral over Fourier
modes of the field with small momenta. As explained above, the quantum
theory with existing continuum limit is a Wick-rotated counterpart
of the statistical physics model describing a second order phase transition.
In the vicinity of a second order phase transition broken and unbroken
symmetry phases are continuously intermixed together, which leads
to the translational invariance of correlation functions of the order
parameter $\phi$. In the case of the first order phase transition,
such invariance is strictly speaking broken in the presence of stochastic
background of nucleating bubbles of the broken symmetry phase, see
the discussion in the previous Section. Therefore, the problem ``at
large'' rewritten in terms of $\phi_{f}$ and $\phi_{s}$ becomes
of Caldeira-Legett type \citep{Caldeira1983}. If we focus our attention
on the physics at scales smaller than the bubble size, translational
invariance does approximately hold, and we can consider $\phi_{s}$
and $\phi_{f}$ as linearly separable (if they are not, we simply
diagonalize the part of the Hamiltonian quadratic in $\phi$).}.

The density matrix $\rho(t,\phi_{s},\phi_{s}')$ of the ``slow''
field or master field configurations is related to the Feynman-Vernon
influence functional $S_{I}[\phi_{1},\phi_{2}]$ of the theory \citep{Calzetta1995}
according to
\[
\rho(t,\phi_{s},\phi_{s}')=\int d\phi_{0}d\phi_{0}^{'}\rho(t,\phi_{0},\phi_{0}')\times
\]
\begin{equation}
\times\int_{\phi_{0}}^{\phi_{s}}d\phi_{1}\int_{\phi_{0}'}^{\phi_{s}'}d\phi_{2}\exp\left(iS[\phi_{1}]-iS[\phi_{2}]+iS_{I}[\phi_{1},\phi_{2}]\right),\label{eq:NRDensityMatrix}
\end{equation}
where
\begin{equation}
S[\phi_{1,2}]=\int d^{d}x\left(\frac{1}{2}(\partial\phi_{1,2})^{2}-\frac{1}{2}m^{2}\phi_{1,2}^{2}-\frac{1}{4}\lambda\phi_{1,2}^{4}\right),\label{eq:EffAction1}
\end{equation}
and

\begin{equation}
S_{I}=-\frac{3}{2}\lambda\int d^{d}x\Delta_{F}(x,x)(\phi_{1}^{2}-\phi_{2}^{2})+\label{eq:FeynmanVernon}
\end{equation}
\[
+\frac{9\lambda^{2}i}{4}\int d^{d}x\,d^{d}y\phi_{1}^{2}(x)(\Delta_{F}(x,y))^{2}\phi_{1}^{2}(y)-
\]
\[
-\frac{9\lambda^{2}i}{2}\int d^{d}x\,d^{d}y\phi_{1}^{2}(x)(\Delta_{-}(x,y))^{2}\phi_{2}^{2}(y)+
\]
\[
\frac{9\lambda^{2}i}{4}\int d^{d}x\,d^{d}y\phi_{2}^{2}(x)(\Delta_{D}(x,y))^{2}\phi_{2}^{2}(y)+\ldots,
\]
where $\phi_{1,2}$ are the Schwinger-Keldysh components of the field
$\phi_{s}$, and $\Delta_{F,-,D}$ are Feynman, negative frequency
Wightman and Dyson propagators of the ``fast'' field $\phi_{f}$,
respectively \footnote{Here, we kept only the leading terms in $\lambda\sim\xi^{4-d}$ as
higher loops as well as other non-renormalizable interactions provide
contributions to the FV functional, which are subdominant (and vanishing!)
in the continuum limit $\xi\to\infty$. }. It is easy to see that the expression (\ref{eq:EffAction1}) is
essentially the same as (\ref{eq:AmplitudeAndFluctuations}), that
is of no surprise since an observer with an IR cutoff cannot distinguish
between $\Phi$ and $\phi_{s}$.

The part of the Feynman-Vernon functional (\ref{eq:FeynmanVernon})
that is interesting for us can be rewritten as
\begin{equation}
S_{I}=i\lambda^{2}\int d^{d}x\,d^{d}y(\phi_{1}^{2}(x)-\phi_{2}^{2}(x))\nu(x-y)(\phi_{1}^{2}(y)-\phi_{2}^{2}(y))-\label{eq:EffActionRepresentation}
\end{equation}
\[
-\lambda^{2}\int d^{d+1}x\,d^{d+1}(\phi_{1}^{2}(x)-\phi_{2}^{2}(x))\mu(x-y)(\phi_{1}^{2}(y)+\phi_{2}^{2}(y))+\ldots
\]
(note that non-trivial effects including the one of decoherence appear
in the earliest only at the second order in $\lambda$). 

An important observation to make is that since the considered non-renormalizable
theory becomes trivial in the continuum limit, see (\ref{eq:Amplitude}),
the kernels $\mu$ and $\nu$ can be approximated as local, i.e.,
$\mu(x-y)\approx\mu_{0}\delta(x-y)$, $\nu(x-y)\approx\nu_{0}\delta(x-y)$.
This is due to the fact that fluctuations $\delta\phi\sim\phi_{f}$
are (almost) decoupled from the master field $\Phi\sim\phi_{s}$ in
the continuum limit, their contribution to (\ref{eq:EffAction1})
is described by the (almost) \emph{Gaussian} functional. Correspondingly,
if one assumes factorization and Gaussianity of the initial conditions
for the modes of the ``fast'' field $\phi_{f}$, the Markovian approximation
is valid for the functional (\ref{eq:EffAction1}), (\ref{eq:FeynmanVernon}).

A rather involved calculation (see \citep{Calzetta1995}) then shows
that the density matrix (\ref{eq:NRDensityMatrix}) is subject to
the master equation 
\begin{equation}
\frac{\partial\rho(t,\phi_{s},\phi_{s}')}{\partial t}=-\int d^{d-1}x[H_{I}(x,\tau),\rho]+\ldots,\label{eq:phi4master}
\end{equation}
\[
H_{I}\approx\frac{1}{2}\lambda^{2}\nu_{0}(\phi_{s}^{2}(\tau,x)-\phi_{s}^{,2}(\tau,x))^{2},
\]
where only terms of the Hamiltonian density $H_{I}$, which lead to
the exponential decay of non-diagonal matrix elements of $\rho$ are
kept explicitly, while $\ldots$ denote oscillatory terms. 

The decoherence time can easily be estimated as follows. If only ``quasi''-homogeneous
master field is kept in (\ref{eq:phi4master}), the density matrix
is subject to the equation
\begin{equation}
\frac{\partial\rho(t,\Phi,\Phi')}{\partial t}=-\frac{1}{2}\lambda^{2}\nu_{0}V_{d-1}[(\Phi-\Phi')^{2}(\Phi+\Phi')^{2},\rho]=-\frac{1}{2}\lambda^{2}\nu_{0}V_{d-1}[(\Phi-\Phi')^{2}\bar{\Phi}^{2},\rho],\label{eq:phi4masterSimplified}
\end{equation}
where $\bar{\Phi}=\frac{1}{2}(\Phi+\Phi')$. We expect that $\bar{\Phi}$
is close to (but does not necesserily coincides with) the minimum
of the potential $V(\Phi)$, which will be denoted $\Phi_{0}$ in
what follows. For $\Phi\approx\Phi'$, i.e., diagonal matrix elements
of the density matrix the decoherence effects are strongly suppressed.
For the matrix elements with $\Phi\ne\Phi'$ the decoherence rate
is determined by 
\begin{equation}
\Gamma=\frac{1}{2}\lambda^{2}\nu_{0}V_{d-1}(\Phi-\Phi')^{2}\bar{\Phi}^{2}\approx\frac{1}{2}\lambda^{2}\nu_{0}V_{d-1}(\Phi-\Phi')^{2}\Phi_{0}^{2}.\label{eq:DecoherenceRate}
\end{equation}
Thus, the decoherence time scale in this regime is 
\begin{equation}
t_{D}\sim\frac{1}{\lambda^{2}\nu_{0}V_{d-1}(\Phi-\Phi')^{2}\Phi_{0}^{2}}.\label{eq:phi4dectime1}
\end{equation}
It is possible to further simplify this expression. First of all,
one notes that $\lambda_{{\rm renorm}}$ will be entering the final
answer instead of the bare coupling $\lambda$. As was discussed above
(and shown in details in \citep{Aizenman1981,Aizenman1982}), the
dimensionless renormalized coupling $g_{{\rm renorm}}$ is suppressed
in the continuum limit as $\frac{{\rm Const.}}{\xi^{d-4}}$, where
$\xi$ is the physical correlation length. Second, the physical volume
$V$ satisfies the relation $V\lesssim\xi^{d-1}$ (amounting to the
statement that the continuum limit corresponds to correlation length
being of the order of the system size). Finally, $\Phi_{0}^{2}\sim\frac{m_{{\rm ren}}^{2}}{\lambda}\sim\xi^{d-6}$,
i.e., every quantity in (\ref{eq:phi4dectime1}) can be presented
in terms of the physical correlation length $\xi$ only. This should
not be surprising. As was argued in the previous Sections, the mean
field theory description holds effectively in the limit $\Lambda\to\infty$
(or $\xi\to\infty$), which is characterized by uncoupling of fluctuations
from the mean field $\Phi$. Self-coupling of fluctuations $\delta\phi$
is also suppressed in the same limit, thus the physical correlation
length $\xi$ becomes a single parameter defining the theory. The
only effect of taking into account next orders in powers of $\lambda$
(or other interactions!) in the effective action (\ref{eq:EffAction1})
and the Feynman-Vernon functional (\ref{eq:FeynmanVernon}) is the
redefinition of $\xi$, which ultimately has to be determined from
observations. In this sense, (\ref{eq:phi4dectime1}) holds to all
orders in $\lambda$, and it can be expected that 
\begin{equation}
t_{D}\gtrsim{\rm Const.}\xi\cdot(\xi/\delta\xi),\label{eq:phi4dectime2}
\end{equation}
where $\delta\xi\sim|\Phi-\Phi'|$ universally for all $\Phi$, $\Phi'$
of physical interest.

According to the expressions (\ref{eq:phi4dectime1}), (\ref{eq:phi4dectime2})
decay of non-diagonal elements of the density matrix $\rho(t,\Phi_{1},\Phi_{2})$
would take much longer than $\xi/c$ (where $c$ is the speed of light)
for $|\Phi_{1}-\Phi_{2}|\ll|\Phi_{1}+\Phi_{2}|$. It still takes about
$\sim\xi/c$ for matrix elements with $|\Phi_{1}-\Phi_{2}|\sim|\Phi_{1}+\Phi_{2}|$
to decay, a very long time in the limit $\xi\to\infty$. 

Finally, if $\bar{\Phi}\ne\Phi_{0}$, i.e., the ``vacuum'' is excited,
$\bar{\Phi}$ returns to minimum after a certain time and fluctuates
near it. It was shown in \citep{Calzetta1995} that the field $\Phi$
is subject to the Langevin equation
\begin{equation}
2\mu_{0}\Phi_{0}\frac{d\Phi}{dt}+m^{2}(\Phi-\Phi_{0})\approx\Phi_{0}\xi(t),\label{eq:LangevinEquationPhi}
\end{equation}
\[
\langle\xi(t)\rangle=0,
\]
\[
\langle\xi(t)\xi(t')\rangle=\nu_{0}\delta(t-t'),
\]
where the random force is due to the interaction between the master
field $\Phi$ and the fast modes $\delta\phi$, determined by the
term $\frac{3}{2}\lambda\Phi^{2}\delta\phi^{2}$ in the effective
action. (The Eq. (\ref{eq:LangevinEquationPhi}) was derived be application
of Hubbard-Stratonovich transformation to the effective action for
the fields $\Phi$ and $\delta\phi$ and assuming that $\Phi$ is
close to $\Phi_{0}$.) The average 
\[
\langle\Phi\rangle-\Phi_{0}\approx(\Phi_{{\rm init}}-\Phi_{0})\exp\left(-\frac{m^{2}}{2\mu_{0}\Phi_{0}}(t-t_{{\rm init}})\right),
\]
so the master field rolling towards the minimum of its potential plays
a role of ``time'' in the theory. The roll towards the minimum $\Phi_{0}$
is very slow, as the rolling time $\sim\frac{\mu_{0}\Phi_{0}}{m^{2}}\sim\frac{\mu_{0}}{\sqrt{\lambda}m}\sim\xi^{d-3}$
is large in the continuum limit $\xi\to0$. Once the field reaches
the minimum, there is no ``time'', as the master field $\Phi$ providing
the function of a clock is minimized. The decoherence would naively
be completely absent for the superposition state of vacua $\pm\Phi_{0}$
as follows from (\ref{eq:DecoherenceRate}). However, the physical
vaccuum as seen by a coarse-grained observer is subject to the Langevin
equation (\ref{eq:LangevinEquationPhi}) even in the closest vicinity
of $\Phi=\pm\Phi_{0}$, and the fluctuations $\langle(\Phi-\Phi_{0})^{2}\rangle$
are never zero; one roughly has
\[
\langle(\Phi-\Phi_{0})^{2}\rangle\sim\frac{\Phi_{0}\nu_{0}}{m\mu_{0}},
\]
which should be substituted in the estimate (\ref{eq:phi4dectime1})
for matrix elements with $\Phi\approx\Phi'\approx\Phi_{0}$. 

What was discussed above holds for coarse-graining scales $p>R_{{\rm IR}}^{-1}$,
where $R_{{\rm IR}}$ is given by the expression (\ref{eq:phi4IRscale}).
If the coarse-graining scale is $p\lesssim R_{{\rm IR}}^{-1}$, the
EFT description breaks down, since at this scale the effective dimensionless
coupling between different modes becomes of the order 1, and the modes
contributing to $\phi_{s}$ and $\phi_{f}$ can no longer be considered
weakly interacting. However, we recall that at probe scales $l>R_{IR}$
the unbroken phase mean field description is perfectly applicable
(see above). This again implies extremely long decoherence time scales. 

The emergent physical picture is the one of entangled states with
coherence surviving during a very long time (at least $\sim\xi/c$)
on spatial scales of the order of at least $\xi$. The largeness of
the correlation length $\xi$ in statistical physics models describing
the vicinity of a first order phase transition implied a large scale
correlation at the spatial scales $\sim\xi$. As was suggested above,
the decoherence is indeed very ineffective in such theories. We shall
see below that the physical picture presented here has a very large
number of analogies in the case of decoherence in quantum gravity.

\subsection{Decoherence in functional Schrodinger picture\label{subsec:Decoherence-in-functional}}

Let us now perform a first quantization analysis of the theory and
see how decoherence emerges in this analysis. As the master field
$\Phi$ is constant in space-time, the field state approximately satisfies
the Schrodinger equation 
\[
\hat{H}_{\Phi}|\Psi(\Phi)\rangle=E_{0}|\Psi(\Phi)\rangle,
\]
where the form of the Hamiltonian $\hat{H}_{\Phi}$ follows straightforwardly
from (\ref{eq:Amplitude}):
\[
\hat{H}_{\Phi}=-\frac{1}{2}V_{d-1}\frac{\partial^{2}}{\partial\Phi^{2}}\pm\frac{1}{2}V_{d-1}m^{2}\Phi^{2}+\frac{1}{4}V_{d-1}\lambda\Phi^{4}.
\]
The physical meaning of $E_{0}$ is the vacuum energy of the scalar
field, which one can safely choose to be $0$. 

Next, one looks for the quasi-classical solution of the Schrodinger
equation of the form $\Psi_{0}(\Phi)\sim\exp(iS_{0}(\Phi))$. The
wave function of fluctuations $\delta\phi$ (or $\phi_{f}$ using
terminology of the previous Subsection) in turn satisfies the Schrodinger
equation
\begin{equation}
i\frac{\partial\psi(\Phi,\phi_{f})}{\partial\tau}=\hat{H}_{\delta\phi}\psi(\Phi,\phi_{f}),\label{eq:FunctionalSchroedinger}
\end{equation}
where $\frac{\partial}{\partial\tau}=\frac{\partial S_{0}}{\partial\Phi}\frac{\partial}{\partial\Phi}$
and $\hat{H}_{\delta\phi}$ is the Hamiltonian of fluctuations $\delta\phi$,
\[
\hat{H}_{\delta\phi}=\int d^{d-1}x\left(-\frac{1}{2}\frac{\partial^{2}}{\partial\delta\phi^{2}}+\frac{1}{2}(\nabla\delta\phi)^{2}+V(\Phi,\delta\phi)\right),
\]
where $V(\Phi,\delta\phi)=\pm\frac{1}{2}m^{2}\delta\phi^{2}+\frac{3}{2}\lambda\Phi^{2}\phi^{2}$,
and the full state of the field is $\Psi(\Phi,\delta\phi)\sim\Psi_{0}(\Phi)\psi(\Phi,\delta\phi)\sim\exp(iS_{0}(\Phi))\psi(\Phi,\delta\phi)$
(again, we naturally assume that the initial state was a factorized
Gaussian). It was previously shown (see \citep{Kiefer1992} and references
therein) that the ``time''-like affine parameter $\tau$ in (\ref{eq:FunctionalSchroedinger})
coincides in fact with the physical time $t$. 

Writing down the expression for the density matrix of the master field
$\Phi$ 
\[
\rho(t,\Phi_{1,}\Phi_{2})={\rm Tr}_{\delta\phi}(\Psi(\Phi_{1},\delta\phi)\Psi^{*}(\Phi_{2},\delta\phi))=
\]
\begin{equation}
=\rho_{0}\int{\cal D}\delta\phi\,\psi(\tau,\Phi_{1},\delta\phi)\psi^{*}(\tau,\Phi_{2},\delta\phi),\label{eq:DensityMatrixFunctionalSchrodinger}
\end{equation}
where
\[
\rho_{0}=\exp\left(iS_{0}(\Phi_{1})-iS_{0}(\Phi_{2})\right),
\]
\[
S_{0}(\Phi)=\frac{1}{2}V_{d-1}(\dot{\Phi})^{2}\mp\frac{1}{2}V_{d-1}m^{2}\Phi^{2}-\frac{1}{4}V_{d-1}\lambda\Phi^{4},
\]
one can then repeat the analysis of \citep{Kiefer1992}. Namely, one
takes a Gaussian ansatz for $\psi(\tau,\Phi,\delta\phi)$ (again,
this is validated by the triviality of the theory)
\[
\psi(\tau,\Phi,\delta\phi)=N(\tau)\exp\left(-\int d^{d-1}p\,\delta\phi(p)\Omega(p,\tau)\delta\phi(p)\right),
\]
where $N$ and $\Omega$ satisfy the equations
\begin{equation}
i\frac{d\log N(\tau)}{d\tau}={\rm Tr}\Omega,\label{eq:Nequation}
\end{equation}
\begin{equation}
-i\frac{\partial\Omega(p,\tau)}{\partial\tau}=-\Omega^{2}(p,\tau)+\omega^{2}(p,\tau),\label{eq:Omegaequation}
\end{equation}
\[
\omega^{2}(p,\tau)=p^{2}+m^{2}+3\lambda\Phi^{2}+\ldots,
\]
and the trace denotes integration over modes with different momenta:
\[
{\rm Tr}\Omega=V_{d-1}\int\frac{d^{d-1}p}{(2\pi)^{d-1}}\Omega(p,\tau).
\]
The expression for $N(t)$ can immediately be found using the Eq.
(\ref{eq:Nequation}) and the normalization condition
\[
\int{\cal D}\delta\phi|\psi(\tau,\Phi,\delta\phi)|^{2}=1
\]
(if $N(\tau)=|N(\tau)|\exp(i\xi(\tau))$, the former completely determines
the absolute value $|N(\tau)|$, while the latter \textemdash{} the
phase $\xi(\tau)$). Then, after taking the Gaussian functional integration
in (\ref{eq:DensityMatrixFunctionalSchrodinger}), the density matrix
can be rewritten in terms of the real part of $\Omega(p,\tau)$ as
\[
\rho(\Phi_{1},\Phi_{2})\approx\frac{\rho_{0}\sqrt{\det({\rm Re}(\Phi_{1}))\det({\rm Re}(\Phi_{2}))}}{\sqrt{{\rm det}(\Omega(\Phi_{1})+\Omega^{*}(\Phi_{2}))}}\exp\left(-i\int^{t}dt'\cdot({\rm Re}\Omega(\Phi_{1})-{\rm Re}\Omega(\Phi_{2}))\right).
\]
Assuming the closeness of $\Phi_{1}$ and $\Phi_{2}$ and following
\citep{Kiefer1992} we expand
\[
\Omega(\Phi_{2})\approx\Omega(\bar{\Phi})+\Omega'(\bar{\Phi})\Delta+\frac{1}{2}\Omega''(\bar{\Phi})\Delta^{2}+\ldots,
\]
where again $\bar{\Phi}=\frac{1}{2}(\Phi_{1}+\Phi_{2})$, $\Delta=\frac{1}{2}(\Phi_{1}-\Phi_{2})$,
and keep terms proportional to $\Delta^{2}$ only. A straightforward
but lengthy calculation shows that the exponentially decaying term
in the density matrix has the form
\begin{equation}
\exp(-D)=\exp\left(-{\rm Tr}\frac{|\Omega'(\bar{\Phi})|^{2}}{({\rm Re}\Omega(\bar{\Phi}))^{2}}\Delta^{2}\right),\label{eq:DecoherenceWidth}
\end{equation}
where $D$ is the decoherence factor, and the decoherence time can
be directly extracted from this expression.

To do so, we note that $\Omega(\Phi)$ is subject to the Eq. (\ref{eq:Omegaequation}).
When $\bar{\Phi}=\Phi_{0}$, one has $\Omega^{2}=\omega^{2}$, and
$\Omega$ does not have any dynamics according to (\ref{eq:Omegaequation}).
However, if $\Phi_{1,2}\ne\bar{\Phi}_{0}$, $\Omega^{2}\ne\omega^{2}$.
As the dynamics of $\Phi$ is slow (see Eq. (\ref{eq:LangevinEquationPhi})),
one can consider $\omega$ as a function of the constant field $\Phi$
and integrate the Eq. (\ref{eq:Omegaequation}) directly. As the time
$t$ enters the solution of this equation only in combination $\omega t$,
one immediately sees that the factor (\ref{eq:DecoherenceWidth})
contains a term $\sim t$ in the exponent, defining the decoherence
time. The latter coincides with the expression (\ref{eq:phi4dectime1})
derived in the previous Section as should have been expected.

Thus, the main conclusion of this Section is that the characteristic
decoherence time scale in non-renormalizable field theories akin to
the $\lambda\phi^{4}$ theory in number of dimensions higher than
$4$ is at least of the order of the physical correlation length $\xi$
of the theory, which is taken to be large in the continuum limit.
Thus, decoherence in the nearly continuum limit is very ineffective
for such theories. 

\section{Decoherence of QFTs on curved space-times\label{sec:Decoherence-in-dS}}

Before proceeding to the discussion of the case of gravity, it is
instructive to consider how the dynamics of decoherence of a QFT changes
once the theory is set on a curved space-time. As we shall see in
a moment, even when the theory is renormalizable (the number of space-time
dimensions $d=d_{{\rm up}}$), the setup features many similarities
with the case of a non-renormalizable field theory in the flat space-time
discussed in the previous Section. 

Consider a scalar QFT with potential $V(\phi)=\frac{1}{2}m^{2}\phi^{2}+\frac{1}{4}\lambda\phi^{4}$
in $4$ curved space-time dimensions. Again, we assume the nearly
critical $"T\to T_{c}"$ case, and that is why the renormalized quadratic
term $\frac{1}{2}m^{2}\phi^{2}$ determining the correlation length
of the theory $\xi\sim m_{{\rm renorm}}^{-1}$ is set to vanish (compared
to the cutoff scale $\Lambda$, again for definiteness $\Lambda\sim M_{P}$). 

The scale $\xi$ is no longer the only relevant one in the theory.
The structure of the Riemann tensor of the space-time (the latter
is assumed to be not too curved) introduces new infrared scales for
the theory, and the dynamics of decoherence in the theory depends
on relation between these scales and the mass scale $m$. Without
a much loss of generality and for the sake of simplicity, one can
consider a $dS_{4}$ space-time characterized by a single such scale
(cosmological constant) related to the Ricci curvature of the background
space-time. It is convenient to write 
\[
V(\phi)=V_{0}+\frac{1}{2}m^{2}\phi^{2}+\frac{1}{4}\lambda\phi^{4},
\]
assuming that the $V_{0}$ term dominates in the energy density.

At spatio-temporal probe scales much smaller than the horizon size
$H_{0}^{-1}\sim\frac{M_{P}}{\sqrt{V_{0}}}$ one can choose the state
of the field to be the Bunch-Davies (or Allen-Mottola) vacuum or an
arbitrary state from the same Fock space. Procedures of renormalization,
construction the effective action of the theory and its Feynman-Vernon
influence function are similar to the ones for QFT in Minkowski space-time.
Thus, so is the dynamics of decoherence due to tracing out unobservable
UV modes; the decoherence time scale is again of the order of the
physical correlation length of the theory:
\[
t_{D}\sim\xi\sim m_{{\rm renorm}}^{-1},
\]
 in complete analogy with the estimate (\ref{eq:phi4dectime2}). This
standard answer is replaced by
\begin{equation}
t_{D}\sim H_{0}^{-1},\label{eq:UV-dSdecoherence}
\end{equation}
when the mass of the field becomes smaller than the Hubble scale,
$m^{2}\ll H_{0}^{2}$, and the naive correlation length $\xi$ exceeds
the horizon size of $dS_{4}$. (The answer (\ref{eq:UV-dSdecoherence})
is correct up to a logarithmic prefactor $\sim\log(H_{0})$.)

It is interesting to analyze the case $m^{2}\ll H_{0}^{2}$ in more
details. The answer (\ref{sec:Decoherence-in-dS}) is only applicable
for a physical observer living inside a single Hubble patch. How does
the decoherence of the field $\phi$ look like from the point of view
of \emph{a meta-observer}, who is able to probe the super-horizon
large scale structure of the field $\phi$ \footnote{This question is not completely meaningless, since a setup is possible
in which the value of $V_{0}$ suddenly jumps to zero, so that the
background space-time becomes Minkowski in the limit $M_{P}\to0$,
and the field structure inside a single Minkowski lightcone becomes
accessible for an observer. If her probe/coarse-graining scale is
$l>H_{0}^{-1}$, this is the question which we are trying to address.}? It is well-known \citep{Starobinsky1986,Starobinsky1994} that the
field $\phi$ in the planar patch of $dS_{4}$ coarse-grained at the
spatio-temporal scale of cosmological horizon $H_{0}^{-1}$ is (approximately)
subject to the Langevin equation
\begin{equation}
3H_{0}\frac{d\phi}{dt}=-m^{2}\phi-\lambda\phi^{3}+f(t),\label{eq:Langevin}
\end{equation}
\[
\langle f(t)f(t')\rangle=\frac{3H_{0}^{4}}{4\pi^{2}}\delta(t-t'),
\]
where average is taken over the Bunch-Davies vacuum, very similar
to (\ref{eq:LangevinEquationPhi}), but with the difference that the
amplitude of the white noise and the dissipation coefficient are correlated
with each other. The corresponding Fokker-Planck equation
\begin{equation}
\frac{\partial P(t,\phi)}{\partial t}=\frac{1}{3H_{0}}\frac{\partial}{\partial\phi}\left(\frac{\partial V}{\partial\phi}P(t,\phi)\right)+\frac{H_{0}^{3}}{8\pi^{2}}\frac{\partial^{2}P}{\partial\phi^{2}}\label{eq:FP}
\end{equation}
describes behavior of the probability $P(t,\phi)$ to measure a given
value of the field $\phi$ at a given moment of time at a given point
of coarse-grained space. Its solution is normalizable and has an asymptotic
behavior 
\begin{equation}
P(t\to\infty,\phi)\sim\frac{1}{V(\phi)}\exp\left(-\frac{8\pi^{2}V(\phi)}{3H_{0}^{4}}\right)\label{eq:FPsolution1}
\end{equation}
As correlation functions of the coarse-grained field $\phi$ are calculated
according to the prescription
\[
\langle\phi^{n}(t,x)\rangle\sim\int d\phi\cdot\phi^{n}P(\phi,t),
\]
(note that two-, three, etc. point functions of $\phi$ are zero,
and only one-point correlation functions are non-trivial) what we
are dealing with in the case (\ref{eq:FPsolution1}) is nothing but
\emph{a mean field theory} with a free energy $F=\frac{8\pi^{2}}{3}V(\phi)$
calculated as an integral of the mean field $\phi$ over the $4-$volume
$\sim H_{0}^{-4}$ of a single Hubble patch. As we have discussed
in the previous Section, decoherence is not experienced as a physical
phenomenon by the meta-observer at all. In fact, the coarse-graining
comoving scale $l_{c}$ separating the two distinctly different regimes
of a weakly coupled theory with a relatively slow decoherence and
a mean field theory with entirely absent decoherence is of the order
\begin{equation}
R_{{\rm IR}}\sim H_{0}^{-1}\exp(S_{dS}),\label{eq:dStransitionscale}
\end{equation}
where $S_{dS}=\frac{\pi M_{P}^{2}}{H_{0}^{2}}$ is the de Sitter entropy
(compare this expression with (\ref{eq:phi4IRscale})). 

\begin{figure}

\includegraphics[scale=0.8]{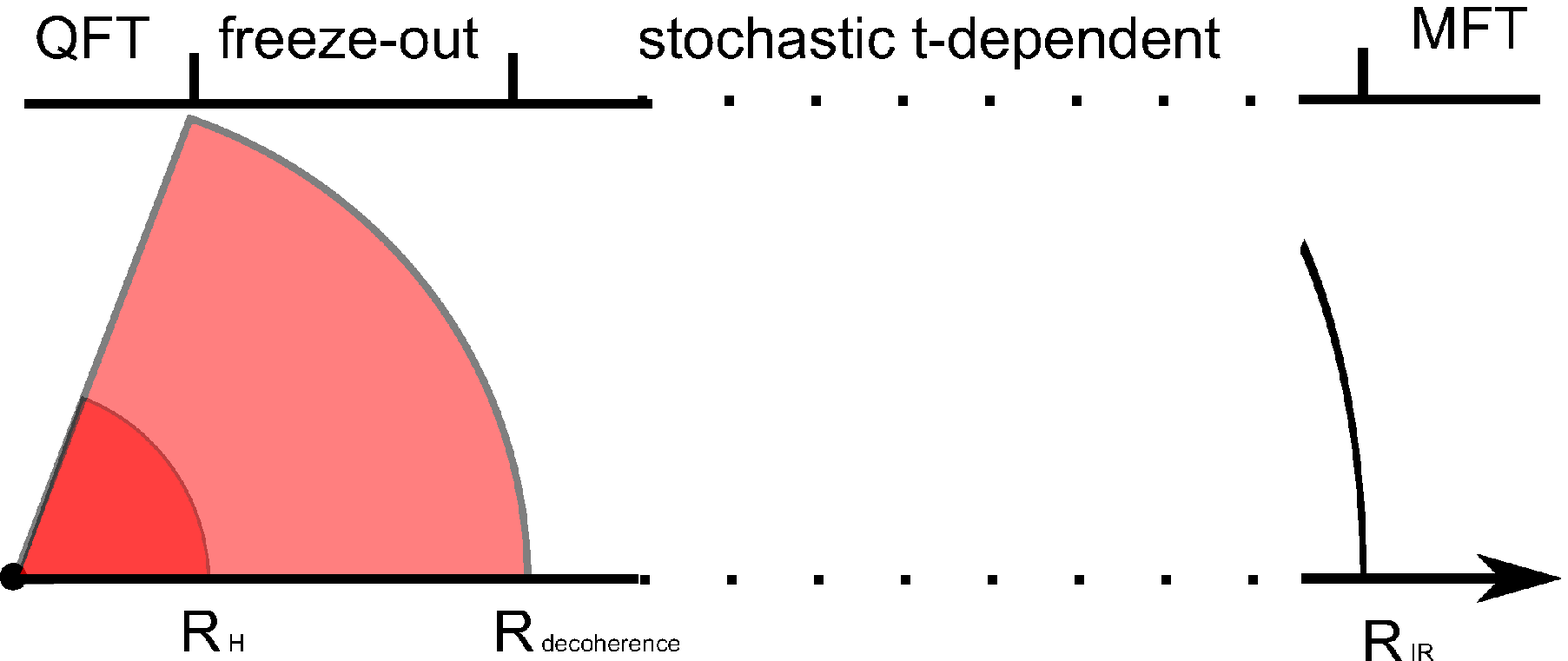}\caption{The hierarchy of decoherence scales for a metaobserver in $dS_{D}$
space. $R_{H}\sim H_{0}^{-1}$ represents the Hubble radius, at comoving
scales $<H_{0}^{-1}$ the correct physical description is the one
in terms of interacting QFT in a de Sitter-invariant vacuum state;
the freeze-out of modes leaving the horizon, vanishing of the decaying
mode and decoherence of the background field (``master field'' $\Phi$)
proceeds at comoving scales $R_{H}<l<R_{{\rm decoherence}}$, where
the latter is by a few efoldings larger than the former, see the next
Section; at $R_{H}<l<R_{{\rm decoherence}}$the field $\Phi$ and
related observables are subject to the Langevin equation (\ref{eq:Langevin})
and represent a stochastic time-dependent background of Hubble patches;
at comoving scales $>R_{{\rm IR}}$ given by (\ref{eq:dStransitionscale})
the stochastic field $\Phi$ reaches the equilibrium solution (\ref{eq:FPsolution1})
of the Fokker-Planck equation (\ref{eq:FP}), and the notion of time
is not well defined; the correct description of the theory is in terms
of the mean/master field with partition function given by (\ref{eq:FPsolution1}).
\label{fig:The-hierarchy}}

\end{figure}

Overall, the physical picture which emerges for the scalar quantum
field theory on $dS_{4}$ background is not very different from the
one realized for the non-renormalizable $\lambda\phi^{4}$ field theory
in Minkowski space-time, see Fig. \ref{fig:The-hierarchy}:
\begin{itemize}
\item for observers with small coarse-graining (comoving) scale $l<H_{0}^{-1}$
the decoherence time scale is at most $H_{0}^{-1}$, which is rather
large physically (of the order of cosmological horizon size for a
given Hubble patch),
\item for a meta-observer with a coarse-graining (comoving) scale $l>R_{{\rm IR}}$,
where $R_{{\rm IR}}$ is given by (\ref{eq:dStransitionscale}), the
decoherence is absent entirely, and the underlying theory is experienced
as a mean field by such meta-observers.
\end{itemize}
Another feature of the present setup which is consistent with the
behavior of a non-renormalizable field theory in a flat space-time
is the breakdown of the effective field theory for the curvature perturbation
in the IR \citep{Arkani-Hamed2007} (as well as IR breakdown of the
perturbation theory on a fixed $dS_{4}$ background) \citep{Woodard2005},
compare with the discussion in Section \ref{sec:Decoherence-non-renormalizable}.
The control on the theory can be recovered if the behavior of observables
in the EFT regime is glued to the IR mean field regime of eternal
inflation \citep{Enqvist2008}.

\section{Decoherence in quantum gravity\label{sec:DecoherenceQG}}

Given the discussions of the previous two Sections, we are finally
ready to muse on the subject of decoherence in quantum gravity, emergence
of time and the cosmological arrow of time, focusing on the case of
$d=3+1$ dimensions. The key observation for us is that the \emph{critical
number of dimensions for gravity is} $d_{{\rm up}}=2$, thus it is
tempting to hypothesize that the case of gravity might have some similarities
with the non-renormalizable theories discussed in Section \ref{sec:Decoherence-non-renormalizable}.

One can perform the analysis of decoherence of quantum gravity following
the strategy represented in Section \ref{subsec:Decoherence-in-EFT},
i.e., studying EFT of the second-quantized gravitational degrees of
freedom, constructing the Feynman-Vernon functional for them and extracting
the characteristic decoherence scales from it (see for example \citep{Barvinsky1998}).
However, it is more convenient to follow the strategy outlined in
Section \ref{subsec:Decoherence-in-functional}. Namely, we would
like to apply the Born-Oppenheimer approximation \citep{Kiefer1992}
to the Wheeler-de Witt equation
\begin{equation}
\hat{H}\Psi=\left(\frac{16\pi G_{ijkl}}{M_{P}^{2}}\frac{\partial^{2}}{\partial h_{ij}\partial h_{kl}}+\sqrt{h^{(3)}}M_{P}^{2}R-\hat{H}_{m}\right)\Psi=0\label{eq:WdW}
\end{equation}
describing behavior of the relevant degrees of freedom (gravity +
a free massive scalar field with mass $m$ and the Hamiltonian $\hat{H}_{m}$).
As usual, gravitational degrees of freedom include functional variables
of the ADM split: scale factor $a$, shift and lapse functions $N_{\mu}$
and the transverse traceless tensor perturbations $h_{ij}$. The WdW
equation (\ref{eq:WdW}) does not contain time at all; similar to
the case of the Fokker-Planck equation (\ref{eq:FP}) for inflation
\citep{Starobinsky1986} the scale factor $a$ replaces it. Time emerges
only after a particular WKB branch of the solution $\Psi$ is picked,
and the WKB piece $\psi(a)\sim\exp(iS_{0})$ of the wave function
$\Psi$ is explicitly separated from the wave functions of the multipoles
$\psi_{n}$ \citep{Kiefer1992}, so that the full state is factorized:
$\Psi=\psi(a)\prod_{n}\psi_{n}$. Similar to the case discussed in
Section \ref{subsec:Decoherence-in-functional}, the latter then satisfy
the functional Schrodinger equations
\begin{equation}
i\frac{\partial\psi_{n}}{\partial\tau}=\hat{H}_{n}\psi_{n},\label{eq:FunctionalSchrodinger}
\end{equation}
(compare to (\ref{eq:FunctionalSchroedinger})). In other words, as
gravity propagates in $d=4>d_{{\rm up}}=2$ space-time dimensions,
we assume a almost complete decoupling of the multipoles $\psi_{n}$
from each other. Their Hamiltonian $\hat{H}_{n}$ is expected to be
Gaussian with possible dependence on $a$: $\psi_{n}$'s are analogous
to the states $\psi$ described by (\ref{eq:FunctionalSchroedinger})
in the case of a non-renormalizable field theory in the flat space-time.
(We note though that this assumption of $\psi_{n}$ decoupling might,
generally speaking, break down in the vicinity of horizons such as
black hole horizons, where the effective dimensionality of space-time
approaches $2$, the critical number of dimensions for gravity.) 

The affine parameter $\tau$ along the WKB trajectory is again defined
according to the prescription 
\[
\frac{\partial}{\partial t}=\frac{\partial}{\partial a}S_{0}\frac{\partial}{\partial a}
\]
and starts to play a role of physical time \citep{Kiefer1992}. One
is motivated to conclude that the emergence of time is related to
the decoherence between different WKB branches of the WdW wave function
$\Psi$, and such emergence can be quantitatively analyzed. 

It was found in \citep{Kiefer1992} by explicit calculation that the
density matrix for the scale factor $a$ behaves as 
\[
\rho(a_{1},a_{2})\sim\exp(-D)
\]
with the decoherence factor for a single WKB branch of the WdW solution
is given by 
\begin{equation}
D\sim\frac{m^{3}}{M_{P}^{3}}(a_{1}+a_{2})(a_{1}-a_{2}).\label{eq:DecoherenceFactor1}
\end{equation}
We note the analogy of this expression with the expression (\ref{eq:DecoherenceWidth})
derived in the the Section \ref{subsec:Decoherence-in-functional}:
decoherence vanishes in the limit $a_{1}=a_{2}$ (or $a_{1}=-a_{2}$)
and is suppressed by powers of cutoff $M_{P}$ ($m/M_{P}$ can roughly
be considered as a dimensionless effective coupling between matter
and gravity). In particular, decoherence is completely absent in the
decoupling limit $M_{P}\to\infty$.

To estimate the involved time scales, let us consider for definiteness
the planar patch of $dS_{4}$ with $a(t)\sim\exp(H_{0}t)$. It immediately
follows from (\ref{eq:DecoherenceFactor1}) that the single WKB branch
decoherence only becomes effective after 
\begin{equation}
H_{0}t_{d}\gtrsim\log\left(\frac{M_{P}^{3}}{m^{3}(a_{1}-a_{2})}\right)\label{eq:QGdecoherence1branch}
\end{equation}
Hubble times, a logarithmically large number of efoldings in the regime
of physical interest, when $M_{P}\gg m_{{\rm phys}}\to0$ (see also
discussion of the decoherence of cosmic fluctuations in \citep{Kiefer2006},
where a similar logarithmic amplification with respect to a single
Hubble time is found). Similarly, the decoherence scale between the
two WKB branches of the WdW solution (corresponding to expansion and
contraction of the inflating space-time)
\[
\psi\sim c_{1}e^{iS_{0}}+c_{2}e^{-iS_{0}}
\]
can be shown to be somewhat smaller \citep{Kiefer1992,Barvinsky1998}:
one finds for the decoherence factor
\[
D\sim\frac{mH_{0}^{2}a^{3}}{M_{P}^{3}},
\]
and the decoherence time (derived from the bound $D(t_{d})\apprge1$)
is given by 
\begin{equation}
H_{0}t_{d}\gtrsim\log\left(\frac{M_{P}^{3}}{mH_{0}^{2}}\right),\label{eq:QGdecoherence2branch}
\end{equation}
still representing a logarithmically large number of efoldings. Taking
for example $m\sim100$ GeV and $H_{0}\sim10^{-42}$ GeV one finds
$H_{0}t_{d}\sim300$. Even for inflaitonary energy scale $H_{0}\sim10^{16}$
GeV the decoherence time scale is given by $H_{0}t_{d}\sim3$ inflationary
efoldings, still a noticeable number. Interestingly, it also takes
a few efoldings for the modes leaving the horizon to freeze and become
quasi-classical. 

Note that (a) $H_{0}$ does not enter the expression (\ref{eq:QGdecoherence1branch})
at all, and it can be expected to hold for other (relatively spatially
homogeneous) backgrounds beyond $dS_{d}$, (b) (\ref{eq:QGdecoherence1branch})
is proportional to powers of effective dimensionless coupling between
matter and gravity, which gets suppressed in the ``continuum''/decoupling
limit by powers of cutoff, (c) decoherence is absent for the elements
of the density matrix with $a_{1}=\pm a_{2}$. These analogies allow
us to expect that a set of conclusions similar to the ones presented
in Sections \ref{sec:Decoherence-non-renormalizable} and \ref{eq:UV-dSdecoherence}
would hold for gravity on other backgrounds as well:
\begin{itemize}
\item we expect the effective field theory description of gravity to break
down in the IR at scales $l\sim l_{{\rm IR}}$ \footnote{Space-like interval $l=\int ds$ connecting two causally unconnected
events.}; the latter is exponentially larger than the characteristic scale
of curvature radius $\sim R_{H}$ of the background; we roughly expect
\begin{equation}
l_{{\rm IR}}\sim R_{H}\exp\left(\frac{{\rm Const.}}{M_{P}R_{H}}\right),\label{eq:QGscale}
\end{equation}
\item at very large probe scales $l>l_{{\rm IR}}$ gravitational decoherence
is absent; a meta-observer testing theory at such scales is dealing
with the ``full'' solution of the Wheeler-de Witt equation, not
containing time in analogy with eternal inflation sale (\ref{eq:dStransitionscale})
in $dS$ space-time filled with a light scalar field,
\item at probe scales $l\lesssim R_{H}$ purely gravitational decoherence
is slow, as it typically takes $t_{D}\gtrsim R_{H}$ for the WdW wave
function $\psi\sim c_{1}\exp(iS_{0}[a])+c_{2}\exp(-iS_{0}[a])$ to
decohere, if time is measured by the clock associated with the matter
degrees of freedom.
\end{itemize}
Finally, it should be noted that gravity differs from non-renormalizable
field theories described in Sections \ref{sec:nonrenorm-QFTs}, \ref{sec:Decoherence-non-renormalizable}
in several respects, two of which might be of relevance for our analysis:
(a) gravity couples to \emph{all} matter degrees of freedom, the fact
which might lead to a suppression of the corresponding effective coupling
entering in the decoherence factor (\ref{eq:DecoherenceFactor1})
and (b) it effectively couples to macroscopic configurations of matter
fields without any screening effects (this fact is responsible for
a rapid decoherence rate calculated in the classic paper \citep{Joos1986}).
Regarding the point (a), it has been previously argued that the actual
scale at which effective field theory for gravity breaks down and
gravity becomes strongly coupled is suppressed by the effective number
of matter fields $N$ (see for example \citep{Dvali2007}, where the
strong coupling scale is estimated to be of the order $M_{P}/\sqrt{N}$,
rather than the Planck mass $M_{P}$). It is in fact rather straightforward
to extend the arguments presented above to the case of $N$ scalar
fields with $Z_{2}$ symmetry. One immediately finds that the time
scale of decoherence between expanding and contracting branches of
the WdW solution is given by
\[
H_{0}t_{d}\gtrsim\log\left(\frac{M_{P}^{3}}{mN^{1/2}H_{0}^{2}}\right)
\]
(to be compared with the Eq. (\ref{eq:QGdecoherence2branch})), while
the single branch decoherence proceeds at time scales of the order
\[
H_{0}t_{d}\apprge\log\left(\frac{M_{P}^{3}}{m^{3}N^{3/2}(a_{1}-a_{2})}\right).
\]
For the decoherence between expanding and contracting WdW branches
discussed in this Section and for the emergence of cosmological arrow
of time, it is important that most of the matter fields are in the
corresponding vacuum states (with the exception of light scalars,
they are not redshifted away), and the effective $N$ remains rather
low, so our estimations remained affected only extremely weakly by
$N$ dependence. As for the point (b), macroscopic configurations
of matter (again, with the exception of light scalars with $m\ll H_{0}$)
do not yet exist at time scales of interest. 

\section{discussion\label{sec:Conclusions}}

We have concluded the previous Section with an observation that quantum
gravitational decoherence responsible for the emergence of the arrow
of time is in fact rather ineffective. If the typical curvature scale
of the space-time is $\sim R$, it takes at least 
\begin{equation}
N\sim\log\left(\frac{M_{P}^{2}}{R}\right)\gg1\label{eq:Nefoldings}
\end{equation}
efoldings for the quasi-classical WdW wave-function $\psi\sim c_{1}\exp(iS_{0}[a])+c_{2}\exp(-iS_{0}[a])$
describing a superposition of expanding and contracting regions to
decohere into separate WKB branches. Whichever matter degrees of freedom
we are dealing with, we expect the estimate (\ref{eq:Nefoldings})
to hold and remain robust.

Once the decoherence happened, the direction of the arrow of time
is given by the vector $\partial_{t}=\partial_{a}S_{0}\partial_{a}$;
at smaller spatio-temporal scales than (\ref{eq:Nefoldings}) the
decoherence factor remains small, and the state of the system represents
a quantum foam, the amplitudes $c_{1,2}$ determining probabilities
to pick an expanding/contracting WKB branch, correspondingly. Interestingly,
the same picture is expected to be reproduced once the probe scale
of an observer becomes larger than characteristic curvature scale
$R$. As we explained above, the ineffectiveness of gravitational
decoherence is directly related to the fact that gravity is a non-renormalizable
theory, which is nearly completely decoupled from the quantum dynamics
of the matter degrees of freedom.

If so, a natural question emerges why do we then experience reality
as a quasi-classical one with the arrow of time strictly directed
from the past to the future and quantum mechanical matter degrees
of freedom decohered at macroscopic scales? Given one has an answer
to the first part of the question, and the quantum gravitational degrees
of freedom are considered as quasi-classical albeit perhaps stochastic
ones, its second part is very easy to answer. Quasi-classical stochastic
gravitational background radiation leads to a decoherence of matter
degrees of freedom at time scale of the order 
\[
t_{D}\sim\left(\frac{M_{P}}{E_{1}-E_{2}}\right)^{2},
\]
where $E_{1,2}$ two rest energies of two quantum states of the considered
configuration of matter (see for example \citep{Blencowe2013,Pikovski2015}).
This decoherence process happens extremely quickly for macroscopic
configurations of total mass much larger than the Planck mass $M_{P}\sim10^{-8}$
kg. Thus, the problem, as was mentioned earlier, is with the first
part of the question.

As there seems to be no physical mechanism in quantized general relativity
leading to quantum gravitational decoherence at spatio-temporal scales
smaller than (\ref{eq:Nefoldings}), an alternative idea would be
to put the burden of fixing the arrow of time on the observer. In
particular, it is tempting to use the idea of \citep{Maccone2009,Lloyd1989},
where it was argued that quasi-classical ${\rm past\to{\rm future}}$
trajectories are associated with the increase of quantum mutual information
between the observer and the observed system and the corresponding
increase of the mutual entanglement entropy. Vice versa, it should
be expected that quasi-classical trajectories ${\rm future\to{\rm past}}$
are associated with the decrease of the quantum mutual information.
Indeed, consider an observer $A$, an observed system $B$ and a reservoir
$R$ such that the state of the combined system $ABR$ is pure, i.e.,
$R$ is a purification space of the system $AB$. It was shown in
\citep{Maccone2009} that 
\begin{equation}
\Delta S(A)+\Delta S(B)-\Delta S(R)-\Delta S(A:B)=0,\label{eq:MutualInfoDifference}
\end{equation}
where $\Delta S(A)=S(\rho_{A},t)-S(\rho_{A},0)$ is the difference
of the von Neumann entropies of the observer subsystem described by
the density matrix $\rho_{A}$, estimated at times $t$ and $0$,
while $\Delta S(A:B)$ is the quantum mutual information difference,
trivially related to the difference in quantum mutual entropy for
subsystems $A$ and $B$. It immediately follows from (\ref{eq:MutualInfoDifference})
that an apparent decrease of the von Neumann entropy $\Delta S(B)<0$
is associated with the decrease in the quantum mutual information
$\Delta S(A:B)<0$, very roughly, erasure of the quantum correlations
between $A$ and $B$ (encoded the memory of the observer $A$ during
observing the evolution of the system $B$). 

As the direction of the arrow of time is associated with the increase
of von Neumann entropy, the observer $A$ is simply unable to recall
behavior of the subsystem $A$ associated with the decrease of its
von Newmann entropy in time. In other words, if the physical processes
representing ``probing the future'' are possible to physically happen,
and our observer is capable to detect them, she will not be able to
store the memory about such processes. Once the quantum trajectory
returns to the starting point (``present''), any memory about observer's
excursion to the future is erased.

It thus becomes clear discussion of the emergence of time (and physics
of decoherence in general) demands somewhat stronger involvement of
an observer than usually accepted in literature. In particular, one
has to prescribe to the observer not only the infrared and ultraviolet
``cutoff'' scales defining which modes of the probed fields should
be regarded as environmental degrees of freedom to be traced out in
the density matrix, but also a quantum memory capacity. In particular,
if the observer does not possess any quantum memory capacity at all,
the accumulation of the mutual information between the observer and
the observed physical system is impossible, and the theorem of \citep{Maccone2009,Lloyd1989}
does not apply: in a sense, the ``brainless'' observer does not
experience time and/or decoherence of any degrees of freedom (as was
earlier suggested in \citep{Lanza2007}).

It should be emphasized that the argument of \citep{Maccone2009}
applies only to quantum mutual information; such processes are possible
that the classical mutual information $S_{cl}(A:B)$ increases, whereas
the quantum mutual information $S(A:B)$ decreases: recall that the
quantum mutual information $S(A:B)$ is the upper bound of $S_{cl}(A:B)$.
Thus, the logic of the expression (\ref{eq:MutualInfoDifference})
applies to observers with ``quantum memory'' with exponential capacity
in the number of qubits\footnote{Number of possible stored patterns is $O(2^{n})$, where $n$ is the
number of qubits in the memory device.} rather than with classical memory with polynomial capacity such as
the ones described by Hopfield networks.

\bibliographystyle{naturemag}
\addcontentsline{toc}{section}{\refname}\bibliography{qgdecoherence4}

\end{document}